\input epsf
\input amssym
\newfam\scrfam
\batchmode\font\tenscr=rsfs10 \errorstopmode
\ifx\tenscr\nullfont
        \message{rsfs script font not available. Replacing with calligraphic.}
        \def\scr{\cal}
\else   
        \font\sevenscr=rsfs7
        \font\fivescr=rsfs5
        \skewchar\tenscr='177 \skewchar\sevenscr='177 \skewchar\fivescr='177
        \textfont\scrfam=\tenscr \scriptfont\scrfam=\sevenscr
        \scriptscriptfont\scrfam=\fivescr
        \def\scr{\fam\scrfam}
        \def\cal{\scr}
\fi
\catcode`\@=11
\newfam\frakfam
\batchmode\font\tenfrak=eufm10 \errorstopmode
\ifx\tenfrak\nullfont
        \message{eufm font not available. Replacing with italic.}
        
\else
	
	\font\sevenfrak=eufm7 \font\fivefrak=eufm5
        
	\textfont\frakfam=\tenfrak
	\scriptfont\frakfam=\sevenfrak \scriptscriptfont\frakfam=\fivefrak
	
\fi
\catcode`\@=\active
\newfam\msbfam
\batchmode\font\twelvemsb=msbm10 scaled\magstep1 \errorstopmode
\ifx\twelvemsb\nullfont\def\Bbb{\bf}
        
	\font\eightbbb=cmb10 at 8pt
	\message{Blackboard bold not available. Replacing with boldface.}
\else   \catcode`\@=11
        \font\tenmsb=msbm10 \font\sevenmsb=msbm7 \font\fivemsb=msbm5
        \textfont\msbfam=\tenmsb
        \scriptfont\msbfam=\sevenmsb \scriptscriptfont\msbfam=\fivemsb
        \def\Bbb{\relax\expandafter\Bbb@}
        \def\Bbb@#1{{\Bbb@@{#1}}}
        \def\Bbb@@#1{\fam\msbfam\relax#1}
        \catcode`\@=\active
	
	\font\eightbbb=msbm8
\fi
        \font\fivemi=cmmi5
        \font\sixmi=cmmi6
        \font\eightrm=cmr8              \def\xrm{\eightrm}
        \font\eightbf=cmbx8             \def\xbf{\eightbf}
        \font\eightit=cmti10 at 8pt     \def\xit{\eightit}
                
        \font\eighttt=cmtt8             
        \font\eightcp=cmcsc8
                      \def\xold{\eighti}
        \font\eightmi=cmmi8
                     \def\xbold{\eightib}
                       \def\old{\teni}
        \font\tencp=cmcsc10

        \font\twelvecp=cmcsc10 scaled\magstep1
        
        \font\sixrm=cmr6
        \font\fiverm=cmr5

        \font\eightsy=cmsy8
        \font\sixsy=cmsy6
        \font\eightsl=cmsl8
        \font\sixbf=cmbx6

	 at10pt	
	\font\twelvehelvbold=phvb at12pt
	 at14pt
	\font\sixteenhelvbold=phvb at16pt

\def\noblackbox{\overfullrule=0pt}
\noblackbox

\def\eightpoint{
\def\rm{\fam0\eightrm}
\textfont0=\eightrm \scriptfont0=\sixrm \scriptscriptfont0=\fiverm
\textfont1=\eightmi  \scriptfont1=\sixmi  \scriptscriptfont1=\fivemi
\textfont2=\eightsy \scriptfont2=\sixsy \scriptscriptfont2=\fivesy
\textfont3=\tenex   \scriptfont3=\tenex \scriptscriptfont3=\tenex
\textfont\itfam=\eightit \def\it{\fam\itfam\eightit}
\textfont\slfam=\eightsl \def\sl{\fam\slfam\eightsl}
\textfont\ttfam=\eighttt \def\tt{\fam\ttfam\eighttt}
\textfont\bffam=\eightbf \scriptfont\bffam=\sixbf 
                         \scriptscriptfont\bffam=\fivebf
                         \def\bf{\fam\bffam\eightbf}
\normalbaselineskip=10pt}

\newtoks\headtext
\headline={\ifnum\pageno=1\hfill\else
	\ifodd\pageno{\eightcp\the\headtext}{ }\dotfill{ }{\old\folio}
	\else{\old\folio}{ }\dotfill{ }{\eightcp\the\headtext}\fi
	\fi}
\def\makeheadline{\vbox to 0pt{\vss\noindent\the\headline\break
\hbox to\hsize{\hfill}}
        \vskip2\baselineskip}
\newcount\infootnote
\infootnote=0
\newcount\footnotecount
\footnotecount=1
\def\foot#1{\infootnote=1
\footnote{${}^{\the\footnotecount}$}{\vtop{\baselineskip=.75\baselineskip
\advance\hsize by
-\parindent{\eightpoint\rm\hskip-\parindent
#1}\hfill\vskip\parskip}}\infootnote=0\global\advance\footnotecount by
1}
\newcount\refcount
\refcount=1
\newwrite\refwrite
\def\oldsize{\ifnum\infootnote=1\xold\else\old\fi}
\def\ref#1#2{
	\def#1{{{\oldsize\the\refcount}}\ifnum\the\refcount=1\immediate\openout\refwrite=\jobname.refs\fi\immediate\write\refwrite{\item{[{\xold\the\refcount}]} 
	#2\hfill\par\vskip-2pt}\xdef#1{{\noexpand\oldsize\the\refcount}}\global\advance\refcount by 1}
	}
\def\refout{\eightpoint\catcode`\@=11
        \xrm\immediate\closeout\refwrite
        \vskip2\baselineskip
        {\noindent\twelvecp References}\hfill\vskip\baselineskip
        \baselineskip=.75\baselineskip
        \input\jobname.refs
        \baselineskip=4\baselineskip \divide\baselineskip by 3
        \catcode`\@=\active\rm}

\def\skipref#1{\hbox to15pt{\phantom{#1}\hfill}\hskip-15pt}

\def\hepth#1{\href{http://xxx.lanl.gov/abs/hep-th/#1}{arXiv:\allowbreak
hep-th\slash{\xold#1}}}

\def\arxiv#1#2{\href{http://arxiv.org/abs/#1.#2}{arXiv:\allowbreak
{\xold#1}.{\xold#2}} [hep-th]} 
 
\def\jhep#1#2#3#4{\href{http://jhep.sissa.it/stdsearch?paper=#2\%28#3\%29#4}{J. High Energy Phys. {\xbold #1#2} ({\xold#3}) {\xold#4}}}

\def\FP#1#2#3{Fortsch. Phys. {\xbold#1} ({\xold#2}) {\xold#3}}

\def\JPA#1#2#3{J. Phys. {\xbf A}{\xbold#1} ({\xold#2}) {\xold#3}}

\def\NPB#1#2#3{Nucl. Phys. {\xbf B}{\xbold#1} ({\xold#2}) {\xold#3}}

\def\PLB#1#2#3{Phys. Lett. {\xbf B}{\xbold#1} ({\xold#2}) {\xold#3}}

\def\PRD#1#2#3{Phys. Rev. {\xbf D}{\xbold#1} ({\xold#2}) {\xold#3}}

\newcount\sectioncount
\sectioncount=0
\def\section#1#2{\global\eqcount=0
	\global\subsectioncount=0
        \global\advance\sectioncount by 1
	\ifnum\sectioncount>1
	        \vskip2\baselineskip
	\fi
\line{\twelvecp\the\sectioncount. #2\hfill}
       \vskip.5\baselineskip\noindent
        \xdef#1{{\old\the\sectioncount}}}
\newcount\subsectioncount
\def\subsection#1#2{\global\advance\subsectioncount by 1
\vskip.75\baselineskip\noindent\line{\tencp\the\sectioncount.\the\subsectioncount. #2\hfill}\nobreak\vskip.4\baselineskip\nobreak\noindent\xdef#1{{\old\the\sectioncount}.{\old\the\subsectioncount}}}
\def\immediatesubsection#1#2{\global\advance\subsectioncount by 1
\vskip-\baselineskip\noindent
\line{\tencp\the\sectioncount.\the\subsectioncount. #2\hfill}
	\vskip.5\baselineskip\noindent
	\xdef#1{{\old\the\sectioncount}.{\old\the\subsectioncount}}}
\newcount\subsubsectioncount
\def\subsubsection#1#2{\global\advance\subsubsectioncount by 1
\vskip.75\baselineskip\noindent\line{\tencp\the\sectioncount.\the\subsectioncount.\the\subsubsectioncount. #2\hfill}\nobreak\vskip.4\baselineskip\nobreak\noindent\xdef#1{{\old\the\sectioncount}.{\old\the\subsectioncount}.{\old\the\subsubsectioncount}}}
\newcount\appendixcount
\appendixcount=0
\def\appendix#1{\global\eqcount=0
        \global\advance\appendixcount by 1
        \vskip2\baselineskip\noindent
        \ifnum\the\appendixcount=1
        \hbox{\twelvecp Appendix A: #1\hfill}\vskip\baselineskip\noindent\fi
    \ifnum\the\appendixcount=2
        \hbox{\twelvecp Appendix B: #1\hfill}\vskip\baselineskip\noindent\fi
    \ifnum\the\appendixcount=3
        \hbox{\twelvecp Appendix C: #1\hfill}\vskip\baselineskip\noindent\fi}
\def\acknowledgements{\vskip2\baselineskip\noindent
        \underbar{\it Acknowledgements:}\ }
\newcount\eqcount
\eqcount=0
\def\Eqn#1{\global\advance\eqcount by 1
\ifnum\the\sectioncount=0
	\xdef#1{{\noexpand\oldsize\the\eqcount}}
	\eqno({\oldstyle\the\eqcount})
\else
        \ifnum\the\appendixcount=0
\xdef#1{{\noexpand\oldsize\the\sectioncount}.{\noexpand\oldsize\the\eqcount}}
                \eqno({\oldstyle\the\sectioncount}.{\oldstyle\the\eqcount})\fi
        \ifnum\the\appendixcount=1
	        \xdef#1{{\noexpand\oldstyle A}.{\noexpand\oldstyle\the\eqcount}}
                \eqno({\oldstyle A}.{\oldstyle\the\eqcount})\fi
        \ifnum\the\appendixcount=2
	        \xdef#1{{\noexpand\oldstyle B}.{\noexpand\oldstyle\the\eqcount}}
                \eqno({\oldstyle B}.{\oldstyle\the\eqcount})\fi
        \ifnum\the\appendixcount=3
	        \xdef#1{{\noexpand\oldstyle C}.{\noexpand\oldstyle\the\eqcount}}
                \eqno({\oldstyle C}.{\oldstyle\the\eqcount})\fi
\fi}
\def\eqn{\global\advance\eqcount by 1
\ifnum\the\sectioncount=0
	\eqno({\oldstyle\the\eqcount})
\else
        \ifnum\the\appendixcount=0
                \eqno({\oldstyle\the\sectioncount}.{\oldstyle\the\eqcount})\fi
        \ifnum\the\appendixcount=1
                \eqno({\oldstyle A}.{\oldstyle\the\eqcount})\fi
        \ifnum\the\appendixcount=2
                \eqno({\oldstyle B}.{\oldstyle\the\eqcount})\fi
        \ifnum\the\appendixcount=3
                \eqno({\oldstyle C}.{\oldstyle\the\eqcount})\fi
\fi}
\def\multi{\global\advance\eqcount by 1}
\def\multieqn#1{({\oldstyle\the\sectioncount}.{\oldstyle\the\eqcount}\hbox{#1})}
\def\multiEqn#1#2{\xdef#1{{\oldstyle\the\sectioncount}.{\old\the\eqcount}#2}
        ({\oldstyle\the\sectioncount}.{\oldstyle\the\eqcount}\hbox{#2})}
\def\multiEqnAll#1{\xdef#1{{\oldstyle\the\sectioncount}.{\old\the\eqcount}}}
\newcount\tablecount
\tablecount=0
\def\Table#1#2{\global\advance\tablecount by 1
       \xdef#1{\the\tablecount}
       \vskip2\parskip
       \centerline{\it Table \the\tablecount: #2}
       \vskip2\parskip}
\newtoks\url
\def\Href#1#2{\catcode`\#=12\url={#1}\catcode`\#=\active#2}
\def\href#1#2{{#2}}

\parskip=3.5pt plus .3pt minus .3pt
\baselineskip=14pt plus .1pt minus .05pt
\lineskip=.5pt plus .05pt minus .05pt
\lineskiplimit=.5pt
\abovedisplayskip=18pt plus 4pt minus 2pt
\belowdisplayskip=\abovedisplayskip
\hsize=14cm
\vsize=19cm
\hoffset=1.5cm
\voffset=1.8cm
\frenchspacing
\footline={}
\raggedbottom

\newskip\origparindent
\origparindent=\parindent

\def\*{\partial}
\def\punkt{\,\,.}
\def\komma{\,\,,}

\def\={\!=\!}
\def\small#1{{\hbox{$#1$}}}

\def\fraction#1{\small{1\over#1}}
\def\fr{\fraction}
\def\Fraction#1#2{\small{#1\over#2}}
\def\Fr{\Fraction}

\def\ie{{\it i.e.}}

\def\id{1\hskip-3.5pt 1}




\def\textfrac#1#2{\raise .45ex\hbox{\the\scriptfont0 #1}\nobreak\hskip-1pt/\hskip-1pt\hbox{\the\scriptfont0 #2}}

\def\LL{{\cal L}}
\def\leftbr{[\![}
\def\rightbr{]\!]}


\def\frac{\Fr}

\def\mathbb{\Bbb}

\def\ZZ{{\Bbb Z}}



\def\LL{{\cal L}}
\def\leftbr{[\![}
\def\rightbr{]\!]}

\def\LL{{\cal L}}
\def\leftbr{[\![}
\def\rightbr{]\!]}


\def\old{\rm}
\def\xold{\xrm}
\def\xbold{\xbf}
\def\oldstyle{}


\ref\Duff{M.J. Duff, {\xit ``Duality rotations in string
theory''}, \NPB{335}{1990}{610}.}

\ref\Tseytlin{A.A.~Tseytlin,
  {\xit ``Duality symmetric closed string theory and interacting
  chiral scalars''}, 
  \NPB{350}{1991}{395}.}

\ref\SiegelI{W.~Siegel,
  {\xit ``Two vierbein formalism for string inspired axionic gravity''},
  \PRD{47}{1993}{5453}
  [\hepth{9302036}].}

\ref\SiegelII{ W.~Siegel,
  {\xit ``Superspace duality in low-energy superstrings''},
  \PRD{48}{1993}{2826}
  [\hepth{9305073}].}

\ref\SiegelIII{W.~Siegel,
  {\xit ``Manifest duality in low-energy superstrings''},
  in Berkeley 1993, Proceedings, Strings '93 353
  [\hepth{9308133}].}

\ref\HullDoubled{C.M. Hull, {\xit ``Doubled geometry and
T-folds''}, \jhep{07}{07}{2007}{080}
[\hepth{0605149}].}

\ref\HullT{C.M. Hull, {\xit ``A geometry for non-geometric string
backgrounds''}, \jhep{05}{10}{2005}{065} [\hepth{0406102}].}

\ref\HullM{C.M. Hull, {\xit ``Generalised geometry for M-theory''},
\jhep{07}{07}{2007}{079} [\hepth{0701203}].}

\ref\HullZwiebachDFT{C. Hull and B. Zwiebach, {\xit ``Double field
theory''}, \jhep{09}{09}{2009}{99} [\arxiv{0904}{4664}].}

\ref\HohmHullZwiebachI{O. Hohm, C.M. Hull and B. Zwiebach, {\xit ``Background
independent action for double field
theory''}, \jhep{10}{07}{2010}{016} [\arxiv{1003}{5027}].}

\ref\HohmHullZwiebachII{O. Hohm, C.M. Hull and B. Zwiebach, {\xit
``Generalized metric formulation of double field theory''},
\jhep{10}{08}{2010}{008} [\arxiv{1006}{4823}].} 

\ref\HohmKwak{O. Hohm and S.K. Kwak, {\xit ``$N=1$ supersymmetric
double field theory''}, \jhep{12}{03}{2012}{080} [\arxiv{1111}{7293}].}

\ref\HohmKwakFrame{O. Hohm and S.K. Kwak, {\xit ``Frame-like geometry
of double field theory''}, \JPA{44}{2011}{085404} [\arxiv{1011}{4101}].}

\ref\JeonLeeParkI{I. Jeon, K. Lee and J.-H. Park, {\xit ``Differential
geometry with a projection: Application to double field theory''},
\jhep{11}{04}{2011}{014} [\arxiv{1011}{1324}].}

\ref\JeonLeeParkII{I. Jeon, K. Lee and J.-H. Park, {\xit ``Stringy
differential geometry, beyond Riemann''}, 
\PRD{84}{2011}{044022} [\arxiv{1105}{6294}].}

\ref\JeonLeeParkIII{I. Jeon, K. Lee and J.-H. Park, {\xit
``Supersymmetric double field theory: stringy reformulation of supergravity''},
\PRD{85}{2012}{081501} [\arxiv{1112}{0069}].}

\ref\HohmZwiebachLarge{O. Hohm and B. Zwiebach, {\xit ``Large gauge
transformations in double field theory''}, \jhep{13}{02}{2013}{075}
[\arxiv{1207}{4198}].} 

\ref\Park{J.-H.~Park,
  {\xit ``Comments on double field theory and diffeomorphisms''},
  \jhep{13}{06}{2013}{098}
  [\arxiv{1304}{5946}].}

\ref\BermanCederwallPerry{D.S. Berman, M. Cederwall and M.J. Perry,
{\xit ``Global aspects of double geometry''}, 
\jhep{14}{09}{2014}{66} [\arxiv{1401}{1311}].}

\ref\CederwallGeometryBehind{M. Cederwall, {\xit ``The geometry behind
double geometry''}, 
\jhep{14}{09}{2014}{70} [\arxiv{1402}{2513}].}

\ref\HohmLustZwiebach{O. Hohm, D. L\"ust and B. Zwiebach, {\xit ``The
spacetime of double field theory: Review, remarks and outlook''},
\FP{61}{2013}{926}
[\arxiv{1309}{2977}].} 

\ref\Papadopoulos{G. Papadopoulos, {\xit ``Seeking the balance:
Patching double and exceptional field theories''}, \arxiv{1402}{2586}.}

\ref\HullGlobal{C.M. Hull, 	
{\xit ``Finite gauge transformations and geometry in double field
theory''}, \arxiv{1406}{7794}.}

\ref\PachecoWaldram{P.P. Pacheco and D. Waldram, {\xit ``M-theory,
exceptional generalised geometry and superpotentials''},
\jhep{08}{09}{2008}{123} [\arxiv{0804}{1362}].}

\ref\Hillmann{C. Hillmann, {\xit ``Generalized $E_{7(7)}$ coset
dynamics and $D=11$ supergravity''}, \jhep{09}{03}{2009}{135}
[\arxiv{0901}{1581}].}

\ref\BermanPerryGen{D.S. Berman and M.J. Perry, {\xit ``Generalised
geometry and M-theory''}, \jhep{11}{06}{2011}{074} [\arxiv{1008}{1763}].}    

\ref\BermanGodazgarPerry{D.S. Berman, H. Godazgar and M.J. Perry,
{\xit ``SO(5,5) duality in M-theory and generalized geometry''},
\PLB{700}{2011}{65} [\arxiv{1103}{5733}].} 

\ref\BermanMusaevPerry{D.S. Berman, E.T. Musaev and M.J. Perry,
{\xit ``Boundary terms in generalized geometry and doubled field theory''},
\PLB{706}{2011}{228} [\arxiv{1110}{97}].} 

\ref\BermanGodazgarGodazgarPerry{D.S. Berman, H. Godazgar, M. Godazgar  
and M.J. Perry,
{\xit ``The local symmetries of M-theory and their formulation in
generalised geometry''}, \jhep{12}{01}{2012}{012}
[\arxiv{1110}{3930}].} 

\ref\BermanGodazgarPerryWest{D.S. Berman, H. Godazgar, M.J. Perry and
P. West,
{\xit ``Duality invariant actions and generalised geometry''}, 
\jhep{12}{02}{2012}{108} [\arxiv{1111}{0459}].} 

\ref\CoimbraStricklandWaldram{A. Coimbra, C. Strickland-Constable and
D. Waldram, {\xit ``$E_{d(d)}\times\hbox{\eightbbb R}^+$ generalised geometry,
connections and M theory'' }, \jhep{14}{02}{2014}{054} [\arxiv{1112}{3989}].} 

\ref\CoimbraStricklandWaldramII{A. Coimbra, C. Strickland-Constable and
D. Waldram, {\xit ``Supergravity as generalised geometry II:
$E_{d(d)}\times\hbox{\eightbbb R}^+$ and M theory''}, 
\jhep{14}{03}{2014}{019} [\arxiv{1212}{1586}].}  

\ref\BermanCederwallKleinschmidtThompson{D.S. Berman, M. Cederwall,
A. Kleinschmidt and D.C. Thompson, {\xit ``The gauge structure of
generalised diffeomorphisms''}, \jhep{13}{01}{2013}{64} [\arxiv{1208}{5884}].}

\ref\ParkSuh{J.-H. Park and Y. Suh, {\xit ``U-geometry: SL(5)''},
\jhep{14}{06}{2014}{102} [\arxiv{1302}{1652}].} 

\ref\CederwallI{M.~Cederwall, J.~Edlund and A.~Karlsson,
  {\xit ``Exceptional geometry and tensor fields''},
  \jhep{13}{07}{2013}{028}
  [\arxiv{1302}{6736}].}

\ref\CederwallII{ M.~Cederwall,
  {\xit ``Non-gravitational exceptional supermultiplets''},
  \jhep{13}{07}{2013}{025}
  [\arxiv{1302}{6737}].}

\ref\SambtlebenHohmI{O.~Hohm and H.~Samtleben,
  {\xit ``Exceptional field theory I: $E_{6(6)}$ covariant form of
  M-theory and type IIB''}, 
  \PRD{89}{2014}{066016} [\arxiv{1312}{0614}].}

\ref\SambtlebenHohmII{O.~Hohm and H.~Samtleben,
  {\xit ``Exceptional field theory II: $E_{7(7)}$''},
  \PRD{89}{2014}{066016} [\arxiv{1312}{4542}].}

\ref\KachruNew{S. Kachru, M.B. Schulz, P.K. Tripathy and S.P. Trivedi,
{\xit ``New supersymmetric string compactifications''}, 
\jhep{03}{03}{2003}{061} [\hepth{0211182}].}

\ref\Condeescu{C. Condeescu, I. Florakis, C. Kounnas and D. L\"ust, 
{\xit ``Gauged supergravities and non-geometric $Q$/$R$-fluxes from
asymmetric orbifold CFT's''}, 
\jhep{13}{10}{2013}{057} [\arxiv{1307}{0999}].}

\ref\CederwallUfoldBranes{M. Cederwall, {\xit ``M-branes on U-folds''},
in proceedings of 7th International Workshop ``Supersymmetries and
Quantum Symmetries'' Dubna, 2007 [\arxiv{0712}{4287}].}



\line{
\epsfysize=18mm
\epsffile{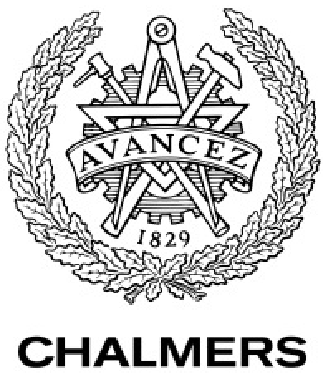}
\hfill}
\vskip-11mm

\line{\hfill Gothenburg preprint}
\line{\hfill September, {\old2014}}

\line{\hrulefill}

\headtext={Cederwall: 
``T-duality and non-geometric solutions...''}

\vfill
\vskip.5cm

\centerline{\sixteenhelvbold
T-duality and non-geometric solutions}

\vskip3mm

\centerline{\sixteenhelvbold
from double geometry}

\vfill

\vfill

\centerline{\twelvehelvbold Martin Cederwall}

\vfill
\vskip-1cm

\centerline{\it Dept. of Fundamental Physics}
\centerline{\it Chalmers University of Technology}
\centerline{\it SE 412 96 Gothenburg, Sweden}

\vfill

{\narrower\noindent \underbar{Abstract:}
Although the introduction of generalised and extended geometry has
been motivated mainly by the appearance of dualities upon reductions on
tori, it has until now been unclear how (all) the duality transformations
arise from first principles in extended geometry. A proposal for
solving this problem is given in the framework of double field theory.
It is based on a clearly defined extension of the definition of
gauge symmetry by isometries of an underlying pseudo-Riemannian
manifold. The ensuing relation between transformations of coordinates
and fields, which is now derived from first principles, differs from
earlier proposals.
\smallskip}
\vfill

\font\xxtt=cmtt6

\vtop{\baselineskip=.6\baselineskip\xxtt
\line{\hrulefill}
\catcode`\@=11
\line{email: martin.cederwall@chalmers.se\hfill}
\catcode`\@=\active
}

\eject

\def\textfrac#1#2{\raise .45ex\hbox{\the\scriptfont0 #1}\nobreak\hskip-1pt/\hskip-1pt\hbox{\the\scriptfont0 #2}}

\def\MM{{\cal M}}

\def\boxit#1{\vbox{\hrule\hbox{\vrule\kern3pt
             \vbox{\kern3pt#1\kern3pt}\kern3pt\vrule}\hrule}}


\section\Introduction{Introduction}The development of extended
geometry has been motivated mainly by the symmetry enhancement in
string theory or M-theory when the models are dimensionally reduced on
tori. The T-duality groups $O(d,d;\ZZ)$ and U-duality groups
$E_{n(n)}(\ZZ)$ are expected to arise as geometric (in a generalised
sense) symmetries of enlarged tori in doubled geometry 
[\Duff\skipref\Tseytlin\skipref\SiegelI\skipref\SiegelII\skipref\SiegelIII\skipref\HullT\skipref\HullDoubled\skipref\HullZwiebachDFT\skipref\HohmHullZwiebachI\skipref\HohmHullZwiebachII\skipref\HohmKwakFrame\skipref\HohmKwak\skipref\JeonLeeParkI\skipref\JeonLeeParkII\skipref\JeonLeeParkIII\skipref\HohmZwiebachLarge\skipref\Park\skipref\BermanCederwallPerry\skipref\CederwallGeometryBehind\skipref\HohmLustZwiebach\skipref\Papadopoulos-\HullGlobal]
and exceptional geometry
[\HullM\skipref\PachecoWaldram\skipref\Hillmann\skipref\BermanPerryGen\skipref\BermanGodazgarPerry\skipref\BermanGodazgarGodazgarPerry\skipref\BermanGodazgarPerryWest\skipref\CoimbraStricklandWaldram\skipref\CoimbraStricklandWaldramII\skipref\BermanCederwallKleinschmidtThompson\skipref\ParkSuh\skipref\CederwallI\skipref\CederwallII\skipref\CederwallUfoldBranes\skipref\SambtlebenHohmI-\SambtlebenHohmII], 
respectively.

Much progress has been made in extended geometry, including results
concerning global questions 
[\HohmZwiebachLarge\skipref\Park\skipref\BermanCederwallPerry\skipref\CederwallGeometryBehind\skipref\HohmLustZwiebach\skipref\Papadopoulos-\HullGlobal].
The complete procedure to obtain the duality groups from first
geometric principles (\ie, as ``generalised isometries'' of certain
backgrounds) has however not been worked out, and it has therefore not
even been clear that the symmetry is present. It has been one of the
main issues in extended geometry how to derive (all) the duality symmetries.
A closely connected and equally urgent 
question is to what extent extended geometry is
able to reproduce non-geometric solutions.
In both cases some educated guesses have been needed for the actual
transformations and holonomies, which have then been constructed ``by hand''.
The purpose of the present paper is to give a geometric framework
for double geometry
where these questions are resolved. 

In Section 2, relevant aspects of double geometry and double
diffeomorphisms are reviewed. Section 3 explains the problem, both in
principle and using a few examples. In Section 4, an appropriate
enlargement of the group of double diffeomorphisms is proposed, and it
is demonstrated how this leads to a resolution of the problems.
Section 5 contains a brief discussion and some outlook.

\section\GeneralisedDiffeomorphisms{Generalised diffeomorphisms in
double field theory}The traditional formulation of generalised
diffeomorphisms relies on 
the introduction of an algebraically invariant tensor of the
continuous version of the duality group. In the case of T-duality and
double field theory, this tensor is an $O(d,d)$-invariant tensor
$\eta_{MN}$.
Generators of infinitesimal generalised diffeomorphisms (acting on
double covectors) then are
$$
\LL_\xi=\xi+a-a^T=L_\xi-a^T\komma
\Eqn\OriginalDiffeos
$$
where $\xi$ is the translational term given by the vector field 
$\xi=\xi^M\*_M$, and $a_M{}^N=\*_M\xi^N$. The transpose is defined by
the invariant metric as $(a^T)_M{}^N=\eta_{MP}\eta^{NQ}a_Q{}^P
=(\eta a^t\eta^{-1})_M{}^N$ ($(\ldots)^t$ denoting ordinary transpose), so
$a-a^T$ is an element of $so(d,d)$.
The commutator of two generalised diffeomorphisms yields
$$
[\LL_\xi,\LL_\chi]=\LL_{\leftbr\xi,\chi\rightbr}\komma\eqn
$$
with $\leftbr\xi,\chi\rightbr=\fr2(\LL_\xi\chi-\LL_\chi\xi)$, provided
that a {\it section condition} holds. This condition is
$$
\eta^{MN}\*_M\otimes\*_N=0\komma\eqn
$$
where the notation indicates that the two derivatives may act on any
field or gauge parameter. A solution to the section condition is
given by the choice of a set of light-like directions on which the
fields do not depend.

The duality group one wants to obtain is $O(d,d;\ZZ)$. In a suitable
(split) basis, where coordinates are $(x^m,\tilde x_m)$ and the
invariant metric has the form
$$
\eta_{MN}=\left[\matrix{0&\id\cr\id&0}\right]\eqn
$$
the section condition is solved by demanding ${\*\over\*\tilde x_m}=0$
on fields and parameters.
This means that 
$$
a-a^T=\left[\matrix{\alpha&\beta\cr0&0}\right]
         -\left[\matrix{0&\beta^t\cr0&\alpha^t}\right]
=\left[\matrix{\alpha&\beta-\beta^t\cr0&-\alpha^t}\right]\punkt
\Eqn\AMinusAT
$$
The exponentiation of such a matrix, is an element 
a$$
J=\left[\matrix{A&B\cr0&(A^{-1})^t}\right]\in O(d,d)\Eqn\RestrictedJ
$$
where $AB^t+BA^t=0$. A generic matrix $J\in O(d,d)$ takes the form
$$
J=\left[\matrix{A&B\cr C&D}\right]\eqn
$$ 
where $AD^t+BC^t=\id$, $AB^t+BA^t=0$ and $CD^t+DC^t=0$.
Matrices with vanishing lower left corner ($C=0$)
belong to $GL(d)\ltimes\Lambda_2\subset O(d,d)$, $\Lambda_2$ being the 
${d(d-1)\over2}$-dimensional module of 2-forms.
They are precisely those that do
not change the solution to the section condition.
Solutions to the section condition are parametrised by a pure spinor
of $O(d,d)$, belonging (up to a real scale) 
to the coset $O(d,d)/(GL(d)\ltimes\Lambda_2)$.

A finite generalised diffeomorphism is of course not given simply by
the exponentiation of eq. (\AMinusAT), but rather leads to the
relation
$$
V'_M(X')=e^{-\xi}(e^{\LL_\xi})_M{}^N V_N(X)\equiv
G_M{}^N(X',X)V_N(X)
\punkt\eqn
$$
Clearly no lower left entry can be contained in the transformation
matrix $G\in O(d,d)$. 

In ref. [\HohmZwiebachLarge], an explicit form of the finite
transformations was conjectured, which reads
$$
\eqalign{
V'_M(X')&=F_M{}^NV_N(X)\komma\cr
F(M)&=\fr2\left[M(M^{-1})^T+(M^{-1})^TM\right]\komma\cr
}\Eqn\HZTransformation
$$
where
$$
M_M{}^N={\*X^N\over\*X'^M}\punkt\Eqn\MDefinition
$$
It was later shown [\Park,\BermanCederwallPerry] that this expression
is correct. The situation deserves some discussion.
It was noted in ref. [\BermanCederwallPerry] that the commutator of
generalised diffeomorphisms may be written in an interesting way as
$$
[\LL_\xi,\LL_\chi]=\LL_{[\xi,\chi]}+\Delta_{\xi,\chi}\punkt
\Eqn\AltComm
$$
The bracket $[\cdot,\cdot]$ is the ordinary Lie bracket of vector
fields, and the ``extra term''
$\Delta_{\xi,\chi}$ has the simple form
$$
\Delta_{\xi,\chi}=-ab^T+ba^T\komma\Eqn\FormOfDelta
$$
with $a_M{}^N=\*_M\xi^N$, $b_M{}^N=\*_M\chi^N$. It is then clear that
the transformation $\Delta_{\xi,\chi}$ itself represents a generalised
diffeomorphism, with parameter
$\zeta^M=-\xi^N(b^T)_N{}^M+\eta^{MN}\*_N\lambda$. The transformation 
$\LL_\zeta=\Delta_{\xi,\chi}$ has no translational term and does not
affect the coordinates. In the basis used above it takes the form
$$
\Delta=\left[\matrix{0&\beta-\beta^t\cr0&0}\right]\punkt\eqn
$$
This element of $O(d,d)$ is nilpotent, and exponentiates to
$e^\Delta=\id+\Delta$. 
Such pure $b$-field transformations form an ideal of the
generalised diffeomorphisms, namely the one preserving
${\*X\over\*X'}=\id$.
This means that given any coordinate transformation matrix $M$, defined by
eq. (\MDefinition), there is an equivalence class $[F(M)]$ of transformation
matrices acting on tensors under generalised diffeomorphisms, and this
class is generated by the representative $F(M)$ of
eq. (\HZTransformation) by $\Delta$-transformations in the ideal. The
matrix $F(M)$ is a canonical representative in the sense that it is
the only element in the class expressible as an algebraic function of
$M$.

The presence of the $\Delta$ transformations leads to an abelian gerbe
structure, described in detail in ref. [\BermanCederwallPerry]:
although the map $M\rightarrow[F(M)]$ is a group homomorphisms (\ie, 
$[F(M)F(N)]=[F(MN)]$), the map $M\rightarrow F(M)$ is not, so that
$$
F(M)F(N)=F(MN)e^{\Delta(M,N)}\punkt\eqn
$$

Since the matrix $M$, due to the section condition, has the form
$$
M=\left[\matrix{m&n\cr0&\id}\right]\komma\eqn
$$
it is straightforward to check that insertion in
eq. (\HZTransformation) for the transformation matrix $F$ also gives a
matrix in $O(d,d)$ of the restricted form (\RestrictedJ), \ie, with
vanishing lower left corner. 

\section\TheProblem{Statement of the problem}The situation described above is
serious if one wants to understand T-duality in 
terms of generalised diffeomorphisms, and if one wants to describe
non-geometric solutions with holonomy in the T-duality group
$O(d,d;\ZZ)$.
The duality group should not be thought of as a global symmetry, {\it
a priori} built into the formalism, but as a generalised isometry
arising for specific solutions, extending the way the mapping class
group $SL(d;\ZZ)$ arises as the discrete isometry group of a torus in
ordinary geometry. The problem is that finite generalised
diffeomorphisms, as described above, do not fill out the whole of
$O(d,d;\ZZ)$. Similarly, if we consider a doubled torus fibered over
some manifold (in the simplest case a circle), the holonomies will
always be elements in $O(d,d;\ZZ)$ of the form (\RestrictedJ),
belonging to the subgroup 
$(GL(d)\ltimes\Lambda_2)(\ZZ)$.

Concerning the question of dualities, there is no need to illustrate
with examples; it is just obvious that some modification is needed if
the full T-duality group is to arise as generalised isometries. This
modification will be given in the following Section.

For the holonomy, I will illustrate with two examples. They 
are to different extents non-geometric, and are
deliberately chosen identical or similar to examples in
ref. [\HohmLustZwiebach], so that the resolution in the following
Section becomes clear.

The first example is obtained by two T-dualities, in the $x$ and
$y$ directions, of a field configuration with $N$ units of
quantised 3-form flux on a
3-torus with radii $R_{1,2,3}$. This gives the field configuration
$$
\eqalign{
ds^2&=f(z)\left({dx^2\over R_1^2}+{dy^2\over
R_2^2}\right)+R_3^2dz^2\komma\cr
b_{12}&=-{Nz\over2\pi R_1^2R_2^2}f(z)\komma\cr
}\Eqn\FirstExample
$$
where the function $f(z)$ is given by
$$
f(z)={1\over1+\bigl({Nz\over2\pi R_1R_2}\bigr)^2}\punkt\eqn
$$
Such non-geometric solutions were constructed in ref. [\KachruNew] 
and discussed
from the perspective of generalised geometry in ref. [\HohmLustZwiebach].
The (inverse) generalised metric $G^{MN}$ for the $x$ and $y$ directions 
encoding this metric and $b$-field is
$$
G^{MN}=\left[\matrix{g^{-1}&-g^{-1}b\cr bg^{-1}&g-bg^{.1}b}\right]
=\left[\matrix{{R_1^2\over f(z)}&0&0&{Nz\over2\pi R_2^2}\cr
                0&{R_2^2\over f(z)}&-{Nz\over2\pi R_1^2}&0\cr
                0&-{Nz\over2\pi R_1^2}&{1\over R_1^2}&0\cr
                {Nz\over2\pi R_2^2}&0&0&{1\over R_2^2}            
}\right]\punkt\eqn
$$ 

The question now is whether this configuration can be constructed in
double geometry. Consider going around the $z$-circle from $z=0$ to
$z=2\pi$. We have
$$
\eqalign{
G^{-1}(0)&=\left[\matrix{R_1^2&0&0&0\cr
0&R_2^2&0&0\cr
0&0&{1\over R_1^2}&0\cr
0&0&0&{1\over R_2^2}}\right]\komma\cr
G^{-1}(2\pi)&=\left[\matrix{R_1^2+{N^2\over R_2^2}&0&0&{N\over R_2^2}\cr
0&R_2^2+{N^2\over R_1^2}&-{N\over R_1^2}&0\cr
0&-{N\over R_1^2}&{1\over R_1^2}&0\cr
{N\over R_2^2}&0&0&{1\over R_2^2}}\right]\punkt\cr
}\eqn
$$
It is easily checked that $G^{-1}(2\pi)=\MM^TG^{-1}(0)\MM$,
where $\MM\in O(d,d;\ZZ)$ is the matrix
$$
\MM=\left[\matrix{
1&0&0&0\cr
0&1&0&0\cr
0&-N&1&0\cr
N&0&0&1
}\right]\punkt\Eqn\FirstExampleHolonomy
$$
This matrix is not of the restricted form discussed above, but instead it
has vanishing upper right corner. After relabeling the two sets of
coordinates $x\leftrightarrow\tilde x$, $\MM$ has the form of
eq. (\RestrictedJ) and can be seen as a generalised
diffeomorphism. Considering fluctuations around this configuration, however,
fields have to depend only on $\tilde x$ and not on $x$, since the
same solution to the section condition has to apply for fields and
gauge transformations.

The second, more genuinely non-geometric, class of configurations has
a holonomy
implying a change $\tau\rightarrow-{1\over\tau}$ of the complex
structure of the 2-torus. It is not constructed to be a true solution
but should be valid as an off-shell configuration. More general such
configurations are considered in refs. [\Condeescu,\HohmLustZwiebach], 
but for our purpose it is enough to let
the $b$-field be zero. Consider any metric on the 2-torus which depends on
$z$ with a period $4\pi$ and has $g(z+2\pi)=g^{-1}(z)$. The holonomy
for the generalised metric is then given by the element
$$
\MM=\left[\matrix{0&\id\cr\id&0}\right]\in O(2,2;\ZZ)
\Eqn\SecondExampleHolonomy
$$ 
Such a T-duality transformation, interchanging $x$ and $\tilde x$,
can not be obtained as a generalised diffeomorphism, and is therefore
not acceptable as holonomy. I will show how this situation may be remedied.

\section\TheProposal{Solution: A geometric proposal for
the gauge symmetries}The main point of the present paper 
is to provide a definition
of gauge transformations that allows for the reinstatement of the 
lower left corner of holonomies and of generalised isometries.
In order to achieve this it will be necessary to find some way of
modifying the concept of gauge transformations, \ie, generalised
diffeomorphisms, so that such elements arise.

Recently, it was shown [\CederwallGeometryBehind] that the invariant
tensor $\eta_{MN}$ may be given a geometric r\^ole. It can be replaced by any
pseudo-Riemannian metric $H_{MN}$ (with split signature), defining a
non-dynamical pseudo-Riemannian background geometry on the doubled manifold. 
The only restriction (which in itself is quite strong) is the
possibility to introduce a section condition. This effectively imposes
a flatness condition in at least half the directions, corresponding to
the $\tilde x$'s.

Derivatives are replaced by covariant derivatives containing the 
torsion-free affine connection compatible with $H$. 
The generators then take the form
$$
\LL_\xi =\xi^MD_M+a-a^T=L_\xi-a^T\komma
\Eqn\NewDiffeos
$$
where now $a_M{}^N=D_M\xi^N$ and $(a^T)_M{}^N=H_{MP}H^{NQ}a_Q{}^P$.
These transformations are generically inequivalent to
eq. (\OriginalDiffeos), but are of course equivalent if $H$ is flat.
The closure of the algebra (more precisely, algebroid) relies on the
first Bianchi identity (the torsion Bianchi identity) together with a
covariant section condition $H^{MN}D_M\otimes D_N=0$, 
and there are no curvature obstructions.
It puts no further restrictions on
curvature than the partial flatness condition already mentioned.
The metric $H$ is a non-dynamical defining background metric
parametrising the embedding $O(d,d)\subset GL(2d)$, and has nothing to
do with the (dynamical) generalised metric in the coset 
$O(d,d)/(O(d)\times O(d))$. 

Even if the definitions (\OriginalDiffeos) and (\NewDiffeos) of the generators
coincide for a flat defining metric $H$, the geometric, rather than
algebraic, interpretation allows for the possibility to let it
transform. Since it is an ordinary metric it will transform as a
tensor of ordinary, not generalised diffeomorphisms (being covariantly
constant, it is invariant
under generalised diffeomorphisms). This may at first sound strange:
we already have a way of changing coordinates through generalised
diffeomorphisms, and it seems contradictory to have two different
types of transformations (and objects behaving like tensors under
both) at once.
This is certainly true for most ordinary diffeomorphisms, but there is
one class which is interesting, namely isometries, leaving the
defining metric, and thus the form of generalised diffeomorphisms 
(\NewDiffeos), unchanged. 

Consider first continuous isometries, generated by $L_u$, where $u$ is
a Killing vector. The observation that the ``covariant'' generalised
diffeomorphisms are formed from $\xi$ in a manner which is manifestly
covariant under ordinary diffeomorphisms leads immediately to
$$
[L_u,\LL_\xi]=\LL_{[u,\xi]}\punkt\eqn
$$
It is important to note that the Killing vector, unlike the parameter
$\xi$, is {\it not}
constrained to obey the section condition, but has arbitrary
coordinate dependence. Such transformations, with
Killing vectors possibly depending on both $x$ and $\tilde x$ may
generate transformations changing the choice of section condition and
having non-vanishing lower left corner.

It is however not continuous isometries that are of most interest, at
least not when the goal is to understand discrete duality
transformations and holonomy. The manifest covariance of eq. (\NewDiffeos)
means that the behaviour of (infinitesimal or finite) generalised
diffeomorphisms under any (finite) isometry is known: they behave like
tensors, and transform with the transformation matrix 
$\MM_M{}^N={\*X^N\over\*X'^M}$ of the
isometry. 
If the transformation matrix $\MM$ is orthogonal (with respect to the
metric $H$), \ie, if $\MM^T\!\MM=\id$, it represents an isometry.
We would again like to stress that, unlike finite generalised
diffeomorphisms, there is no need for parameters to satisfy the
section condition.
It can be
noted that also anti-isometries, with $H\rightarrow-H$ (which is
meaningful on a manifold with split signature), are allowed,
since $H$ appears only quadratically in $\LL_\xi$.

The group of isometries ${\cal I}$ acts as automorphisms of the original
algebra of generalised diffeomorphisms. Some of them may be inner
automorphisms, or 
at least uninteresting, if they correspond to coordinate
transformations already realisable as finite generalised
diffeomorphisms. Some, on the other hand, may act as {\it outer}
automorphisms. This applies to any isometry that rotates the solution
of the section condition. In ordinary geometry, all isometries
act as inner automorphisms, but in doubled geometry, the section condition
prevents the generalised diffeomorphisms to rotate its solution. The
elements in ${\cal I}$ that rotate the solution to the 
section condition (having
entries in the lower left corner) give a geometric tool to realise 
such transformations, especially in the case of a flat torus, where
${\cal I}$ contains the group
$$
SL(2d;\ZZ)\cap O(d,d)=O(d,d;\ZZ)\komma\eqn
$$
which is the isometry subgroup of the mapping class group.

The proposal for a better definition of the gauge transformations should 
now be obvious. It reads:


\item{$\bullet$}{\it The gauge transformations are
declared to be generalised 
diffeomorphisms, together with their automorphisms.}

By elevating the status of isometries, 
\ie, automorphisms
of generalised diffeomorphisms, to gauge symmetries, there are no
longer any outer automorphisms. They are used in order to obtain the
full duality group, and the full holonomy. On a flat torus, this
definition enlarges the gauge symmetries by homotopically non-trivial
sectors which were inaccessible in the earlier formulation (but
nevertheless sometimes inserted ``by hand'').

Take an element $\MM\in{\cal I}$. As shown above, covectors
transform as $V\rightarrow\MM V$. A transformation matrix changes
according to $M\rightarrow\MM M\MM^{-1}=\MM M\MM^T$, which is
consistent with 
$$
\eqalign{
F(M)&=\fr2\left[M(M^{-1})^T+(M^{-1})^TM\right]\cr
&\rightarrow
\fr2\left[\MM M\MM^T((\MM M\MM^T)^{-1})^T
                +((\MM M\MM^T)^{-1})^T\MM M\MM^T\right]\cr
&=\MM F(M)\MM^T\cr
}\eqn
$$
(this is of course just another way of stating the automorphism property).
A consequence of the extended definition of gauge transformations is
that the transformation matrix associated with an isometry is $\MM$
itself, and not the formal extension $F(\MM)$ of the relation
(\HZTransformation) to arbitrary elements in $GL(2d)$ or $O(d,d)$. 
In earlier work [\HohmLustZwiebach], the latter
has been used, leading to $F(\MM)=\MM^2$, and in order to guess a
coordinate transformation corresponding to a certain holonomy or
T-duality transformation, a square root of the transformation matrix
was needed. This lead to some peculiar behaviour, which in some cases, like
our second example above, included orientation-reversing
transformations. 
This is not the case in the present proposal.

Let us now go back to the problem and to the examples.

Concerning duality transformations in $O(d,d;\ZZ)$, they are now all
present as gauge 
transformations in a situation with a flat defining metric.

In the first example of a non-geometric field configuration, given by
eq. (\FirstExample), it was seen that the problem in a sense could be
circumvented by a reinterpretation of the physical subspace (the
choice of solutions to the section condition). With the new proposal,
this is no longer necessary. The holonomy (\FirstExampleHolonomy) is a
discrete gauge transformation disregarding the choice of section
condition. The situation is improved, in the sense that fluctuations
around this configuration are no longer restricted to depend only on
$\tilde x$'s, but can obey any solution to the section condition,
including dependence on $x$'s only (in this patch). Of course, when
moving to other patches, the
section condition transforms accordingly.

In the second example with holonomy given by
eq. (\SecondExampleHolonomy), 
and any other with generic enough holonomy,
the only possibility with the previous
understanding was to put in the desired holonomy by hand. 
In the present picture, there is a geometric prescription for the
transformation which makes it a gauge transformation. In the example,
the actual coordinate transformation is the exchange of $x$ and
$\tilde x$, not its square root.
Again, the fields in a certain patch must obey some section condition,
which then is rotated by the holonomy.

\section\Conclusions{Conclusions}It has been demonstrated how the
gauge symmetries of double field theory may be extended to include
isometries of the non-dynamical defining metric. They act as
automorphisms of the generalised diffeomorphisms, and the important part
consists of the outer automorphisms. An analogous statement should be
true for exceptional geometry, \ie, extended geometry in the context
of U-duality, based on exceptional groups. In those cases, the
geometric interpretation of the invariant tensors, corresponding to
the results of ref. [\CederwallGeometryBehind], has not yet been
worked out (the interpretation of the invariant tensor as a metric
tensor is particular to doubled geometry).

Since the section condition, with the earlier definition of gauge
symmetries, has been responsible for the restriction of the
transformation matrices, a belief has sometimes been expressed that
recovery of the full duality group might require a relaxation of the
section condition. This turned out not to be the case. However, it is
clear that the present construction requires the {\it solution} to the
section condition to be multiple valued (for non-geometric field
configurations), and subject to 
transformations including the holonomy described earlier.
 
An possible objection to the present construction is that it provides
more of a patching up of the problematic situation than a radical
solution. Although the construction has a firm geometric
basis, this may be so, and if there is a better one, it is likely
to arise in a context where the section condition is abandoned
altogether, and instead arises dynamically. 
Suppose for example that a solution to the section condition is
imposed by the choice of a pure spinor. These variables must then
locally be pure gauge degrees of freedom.
If such a formulation
exists, it should probably contain (ordinary) general coordinate
invariance as an essential ingredient. 

\acknowledgements{The author wishes to thank David Berman, Jeong-Hyuck
Park, Daniel Waldram, Lisa Carbone, and especially Olaf Hohm, for discussions.}

\refout
\end